\documentclass[aps,prl,twocolumn,preprintnumbers]{revtex4}

\usepackage{slashed}
\usepackage[sumlimits,intlimits,namelimits]{amsmath} 
\usepackage[english]{babel}
\usepackage{graphicx}

\newcommand{\ice}[1]{\relax}

\begin{document}

\title{Inclusive semileptonic \boldmath $B$ \unboldmath decays from QCD \\ 
with NLO accuracy for power suppressed terms
}
\author{Thomas Mannel$^1$, Alexei A. Pivovarov$^{1,2}$, Denis Rosenthal$^1$}
\affiliation{$^1$ Theoretische Physik 1,
Universit\"at Siegen, D-57068 Siegen, Germany\\
  $^2$ Institute for Nuclear Research of the
  Russian Academy of Sciences, Moscow 117312}
\preprint{SI-HEP-2014-04}
\preprint{QFET-2014-06}
\begin{abstract}\noindent
We present the results of a calculation of the perturbative QCD corrections 
for the semileptonic inclusive width of a heavy flavored meson.  
Within the Heavy Quark Expansion we analytically compute the QCD correction 
to the coefficient of power suppressed 
contribution of the chromo-magnetic operator in the limit of vanishing mass
of the final state quark.
The important phenomenological applications are decays of bottom
mesons, and to the less extend, charmed mesons. 
\end{abstract}

\maketitle

\section{Introduction}
\noindent
With the success of the LHC mission and the Higgs boson discovery 
the validity of Standard Model~(SM) 
as the theory of particle interactions at  energies
below $1~{\rm TeV}$ has been convincingly proven. 
However, it is hard to expect
that we shall be able to explore still higher energy regions
in the same manner, namely by a direct observation of new physics phenomena.
It is conceivable that new
phenomena beyond the SM can only be identified through 
detecting slight discrepancies between
theoretical predictions within the SM
and precision  measurements 
at low energy with available machines.
For this program to succeed, accurate theoretical predictions
withing the SM are of crucial importance, especially precise 
numerical values of key parameters of the SM are necessary.

In this respect, there are a few places that provide valuable information.
The muon decay is important for the
determination of the Fermi constant $G_F$ with high 
accuracy~\cite{kinoshita,berman}. 
To match the precision of the experimental data in this case,  
the theoretical calculations have to be performed with very high
accuracy.
In this case this is feasible,  
since the purely leptonic decays are well described with
perturbation theory and the expansion parameter 
$\alpha\approx 1/137$ is small.
The latest theoretical result includes the  
second order (NNLO) radiative correction
in the fine structure
constant expansion~\cite{vanRitbergen:1998yd}
\begin{equation} \label{rate-muom1}
\Gamma(\mu\to\nu_\mu e\bar{\nu}_e)/\hat\Gamma^0
=1+\left(\frac{25}{8}-\frac{\pi^2}{2}\right)
\frac{\alpha}{\pi}+6.74\left(\frac{\alpha}{\pi}\right)^2 \nonumber
\end{equation} 
with $\hat\Gamma^0=G_F^2 m_\mu^5/(192 \pi^3)$ and $m_\mu$ is the muon mass.
It results in an ${\cal O}$(1ppm) accuracy of theoretical expression
that is competitive for comparison with  
experimental data.

There is a common belief that the flavor physics of quarks  
is one of the most promising places for search of new
physics~\cite{flavor-review}.
The relevant SM parameters in this sector are the Fermi constant and 
quark mixing
parameters gathered in the CKM matrix.
While the quark weak decays are mediated through charged currents at 
tree level (which are believed not to have sizable contributions 
of possible 
new physics), their study is of
paramount importance for precise determination of the numerical values
of the CKM matrix elements.
In contrast to leptons, obtaining 
a theoretical prediction for processes with quarks 
requires the use of genuinely
nonperturbative computational methods (like QCD lattice calculations)
due to confinement. 
Nevertheless, for heavy mesons the theoretical treatment is somewhat 
easier 
because the large mass of the heavy quark opens the possibility for an 
expansion in powers of $\Lambda/m_Q$ where $m_Q$ is the quark 
mass and $\Lambda \sim 500~{\rm MeV}$ is a hadronic 
scale~\cite{hqsym}. Top quarks do not form mesons 
due to the short top quark lifetime, 
charmed mesons are probably not heavy
enough, rendering the application of the Heavy-Quark Expansion (HQE) marginal,
but the case of bottom-meson decays is 
certainly tractable in this way and thus has been
intensively studied.
The technique is applicable to 
$b \to u$ and $b\to c$ transition and both to semileptonic and purely
hadronic decays. For definiteness, we will stick to
semileptonic $b \to c $ 
decays.

Over the last ten years the HQE in inclusive
semileptonic $b \to c $ 
decays
has been 
refined to such an extend that the remaining theoretical uncertainty
in 
the prediction of the total inclusive 
rate for $B \to X_c \ell \bar{\nu}$ has reached a level of less than 
two percent. 
The structure of the HQE in the case at hand is given by~\cite{mannel-bigi}
\begin{eqnarray} \label{rate}
&&\Gamma (B \to X_c \ell \bar{\nu}_\ell)/\Gamma^0 =
|V_{cb}|^2 \left[ a_0(1 + \frac{\mu_\pi^2}{2m_b^2}) \right.   \\
&& \qquad \qquad +  a_2 \frac{\mu_G^2}{m_b^2}  
 \left. + a_3\frac{\bar{\rho}^3}{m_b^3} 
+ {\cal O}\left( \frac{\Lambda^4}{m_b^4}\right)\right]  \nonumber
\end{eqnarray} 
where 
$\Gamma^0=G_F^2 m_b^5/(192 \pi^3)$,  $m_b$ is the $b$-quark mass,    
$\mu_\pi$ (the kinetic energy parameter), 
$\mu_G$ (the chromo-magnetic parameter), and 
$\bar\rho$ are nonperturbative contributions with numerical 
values of the order of $\Lambda$. 
The coefficients $a_i$ are functions of 
the quark (and, in general, lepton) 
masses and have a perturbative expansion in the 
strong coupling constant $\alpha_s (m_b)$. 
The leading term $a_0$ is known  analytically
to~${\cal O} (\alpha_s^2)$ 
precision in the massless limit of the final state quark~\cite{Ritbergen2}.
At NNLO the mass corrections have been  analytically accounted 
for as an expansion in ref.~\cite{Pak:2008qt} 
and numerically 
in~\cite{Melnikov:2008qs}. 
The coefficient of the 
kinetic energy parameter is linked to $a_0$ by Lorentz invariance, 
see the explicit analysis 
in~\cite{becher}. 
The parametrically largest contribution to the width currently unknown 
is the $\alpha_s$ correction to the coefficient of the 
chromo-magnetic parameter $a_2$, 
which has been investigated recently 
in~\cite{Gambino}, where a numerical result for this contribution 
has been obtained. From the numerical study
performed 
in~\cite{Gambino} one can infer that 
the $\alpha_s$ corrections to $a_2$ are of the expected size.  

In this letter we report on an analytical calculation of
corrections to $a_2$
in the limit of vanishing charmed quark mass. As it turns out, 
the precision gained in this approximation is sufficient
for phenomenological applications.

\section{Outline of the Calculation}
\noindent
The rate~(\ref{rate}) is obtained from taking the absorptive part of 
the forward matrix element of the transition operator
$T$~\cite{width-trans},
\begin{equation}\label{eq:trans-op}
T = i\int d^4 x \,   
T\left[ H_{\rm eff} (x) H_{\rm eff} (0) \right] 
\end{equation} 
where $H_{\rm eff}$ is the effective Hamiltonian for the semileptonic 
transition 
\begin{equation}
\label{operators}
H_{\rm eff} = 2\sqrt{2}G_FV_{cb}(\bar{b}_L \gamma_\mu c_L) 
(\bar{\nu}_L \gamma^\mu \ell_L) .
\end{equation} 
In order to make the dependence of the width on the
heavy quark mass  $m_b$ explicit and to build up 
an expansion in $\Lambda/m_b$,  
one matches 
a time-ordered product of full QCD operators $H_{\rm eff}$ 
in~(\ref{operators}) 
on an expansion in terms of Heavy Quark Effective Theory 
(HQET)~\cite{mannelRyzhak,manohar1}
\begin{equation}
\label{HQE-1}
({\rm Im} \, T)/R_0=
C_0 {\cal O}_0  + C_v\frac{{\cal O}_v}{m_b}
+ C_\pi \frac{{\cal O}_\pi}{2m_b^2}  
+ C_G\frac{{\cal O}_G}{2m_b^2}
\end{equation}
where
$R_0=\pi\Gamma_0|V_{cb}|^2$. The local operators~${\cal O}_i$
in the 
expansion~(\ref{HQE-1}) are
ordered by their dimensionality
${\cal O}_0 =\bar{h}_v h_v $,
${\cal O}_v =\bar{h}_v v\pi h_v $, ${\cal O}_\pi
=\bar{h}_v\pi_\perp^2 h_v$,
${\cal O}_G=\bar{h}_v\frac{1}{2}[\slashed{\pi}_\perp,\slashed{\pi}_\perp] h_v$.
Here
$v$ is the velocity of the heavy hadron appearing in the HQET 
construction,
$\pi_\mu =i\partial_\mu+g_sA_\mu $ 
is the covariant 
derivative of QCD, $\pi^\mu =v^\mu (v\pi)+\pi^\mu_\perp$, and 
$h_v$ is the heavy-quark
field  entering the HQET 
Lagrangian~\cite{mannelRyzhak,manohar1}.
The expansion~(\ref{HQE-1}) is a matching relation
from QCD to HQET with
proper operators up to dimension five with the corresponding 
coefficient functions. 
Note that the operator 
${\cal O}_v$
will be eliminated by using the equation of motion for 
$h_v$ once the forward matrix elements with meson
states are taken.
The Lagrangian for the modes $h_v$ is given by 
\begin{equation}
{\cal L}={\cal O}_v+\frac{1}{2m_b}({\cal O}_\pi+C_m(\mu){\cal O}_G)
+O\left(\frac{\Lambda^2}{m_b^2}\right)
\end{equation}
with
\begin{equation}
C_m(\mu)=1+\frac{\alpha_s(\mu)}{2\pi}
\left\{C_F+C_A\left(1+
\ln\frac{\mu}{m_b}\right)\right\}
\end{equation}
being the coefficient of the chromo-magnetic operator ${\cal O}_G$
including the ${\cal O} (\alpha_s)$ QCD correction~\cite{chromo-anom}.
Note that we define the modes $h_v$ such that terms 
of the order $O(1/m_b^2)$ in the Lagrangian 
contain no time derivative~\cite{manohar1,koerner}.

It is convenient to choose the local operator  
$\bar{b}\slashed{v} b$ (defined in full QCD) as a leading term of 
heavy quark expansion~\cite{manohar2}.
Indeed, the current $\bar{b}\gamma_\mu b$  is 
conserved and thus its forward matrix element with
hadronic states is 
absolutely normalized.
For implementing this 
one needs an expansion (matching) of 
a full QCD local operator $\bar{b}\slashed{v} b$ in
HQE through HQET operators. The expansion reads
\begin{equation}\label{eq:local-bvb}
\bar{b}\slashed{v} b
={\cal O}_0+{\tilde C}_\pi\frac{{\cal O}_\pi}{2m_b^2}
+ {\tilde C}_G\frac{{\cal O}_G}{2m_b^2}
+O(1/m_b^3)
\end{equation}
and is valid including the radiative corrections of 
order $\alpha_s$. Thus, the  
leading power operator has no corrections and the kinetic operator has
the same coefficient as the leading one due to Lorentz  invariance.

Substituting expansion~(\ref{eq:local-bvb})
into~(\ref{HQE-1}) one obtains after using the equation
of motion for the operator ${\cal O}_v$  in the forward 
matrix elements
\begin{eqnarray}
({\rm Im} \, T)/R_0&=&C_0 \left\{\bar{b}\slashed{v} b  
 -\frac{{\cal O}_\pi}{2m_b^2}\right\}\nonumber \\
&&+ \left\{-C_vC_m+C_G - {\tilde C}_GC_0\right\}
\frac{ {\cal O}_G}{2m_b^2}.
\label{HQE}
\end{eqnarray}  
The numerical value for the chromo-magnetic moment parameter  
$\mu_G^2 $ related to the forward matrix element of the operator
${\cal O}_G$ is usually 
taken from the mass splitting between the pseudoscalar and 
vector ground-state mesons. 
The mass difference of bottom mesons
$m_{B^*}^2-m_{B}^2=\Delta m_B^2 =0.49~{\rm GeV}^2$ 
is given by 
\begin{equation} \label{delM}
\frac{1}{2 M_B }C_m(\mu)\langle B(p_B)| {\cal O}_G|B(p_B)\rangle=
\frac{3}{4}\Delta m_B^2
\end{equation} 
where we use the usual relativistic normalization of the states. 

\begin{figure}
\begin{center}
\includegraphics[width=0.21\textwidth]{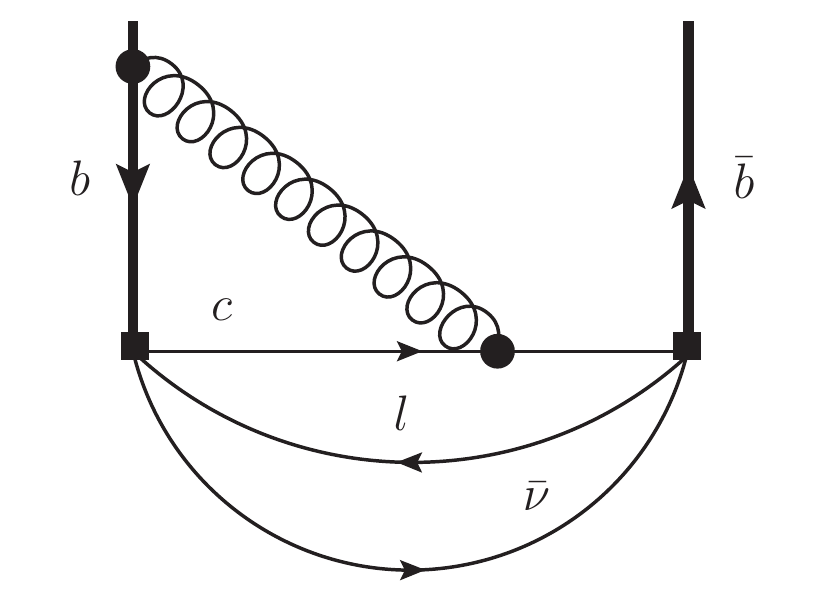}
\includegraphics[width=0.21\textwidth]{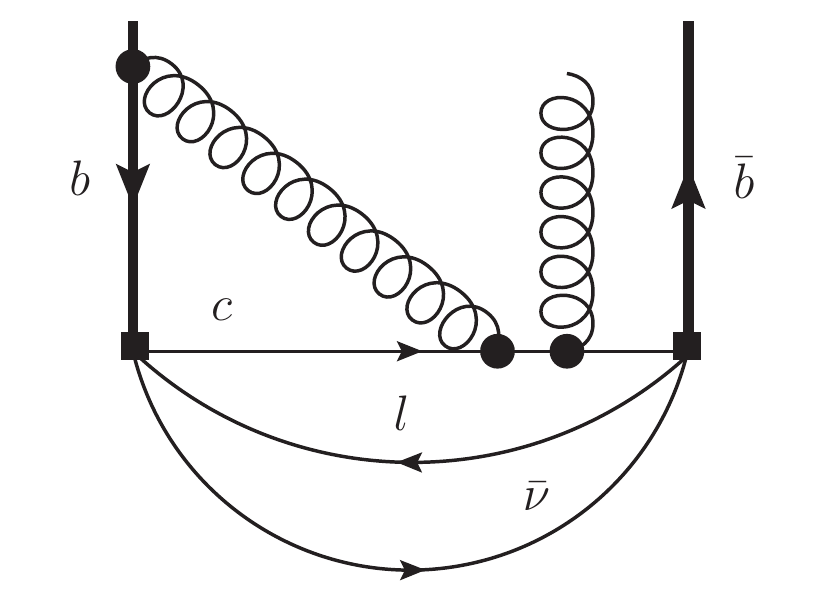}
\end{center}
\vspace{-0.75cm}
\caption{Perturbation theory diagrams for the 
matching computation, 
(left) - width type, right - power correction 
type (in an external gluon field)}
\label{fig:diags-1}
\vspace{-0.5cm}
\end{figure}

Taking the forward matrix element of (\ref{HQE}) one gets
\begin{eqnarray}
&&\Gamma(B\to X_c\nu\ell)
=\Gamma_0|V_{cb}|^2\left\{
C_0 \left(1+\frac{\mu_\pi^2}{2 m_b^2}\right)\right.\nonumber \\ 
\label{HQEfin}
&& \qquad \quad \left.+ \left(-C_v+\frac{C_G - {\tilde C}_G C_0}{C_m}\right)
\frac{3\Delta m_B^2}{8m_b^2}\right\} \, . 
\end{eqnarray} 
The matching procedure is straightforward and consists in computing
matrix elements with partonic states (quarks and gluons on shell)
at both sides of the expansion~(\ref{HQE-1}).
In this way the coefficient function $C_0$
of the dimension three operator $\bar{h}_v h_v$
determines the total width of the heavy quark and at the same
time the leading contribution to the width of a bottom hadron. 
Going to order $\alpha_s$, the calculation of the 
transition operator $T$ in~(\ref{eq:trans-op})
requires to consider three-loop diagrams with external heavy quark
lines on shell.
The leading order 
result is well known and requires the
calculation of the two-loop Feynman integrals of the simplest topology
-- the sunset type ones~\cite{we-annals}.
At the NLO level 
one needs the on-shell tree-loop integrals with massive 
lines.
The computation has been performed in dimensional regularization used for
both ultraviolet and infrared singularities. We used 
the systems of symbolic
manipulations REDUCE~\cite{reduce} and Mathematica~\cite{Mathe}
with original codes written for the calculation. 
The reduction to master integrals
has been done within the integration by parts 
technique~\cite{ibp-tech}. The original codes have been used for most
of the diagrams and then the program LiteRed~\cite{Lite} 
has been used for checking
and further application to complicated vertex diagrams. 
The master integrals have been computed directly and then checked
with the program HypExp~\cite{hypexp}.
The renormalization is performed on-shell by the multiplication of the
bare (direct from diagrams) results by the renormalization constant 
$Z_2^{OS}$
\begin{equation} \label{ZOS}
Z_2^{OS}=1-C_F\frac{\alpha_s}{4\pi}\left(\frac{3}{\epsilon}
+3\ln\left(\frac{\mu^2}{m_b^2}\right)+4\right) .
\end{equation}
In fig.~\ref{fig:diags-1} we show some typical three loop diagrams. 
By using the described 
methods one reproduces the known result~\cite{becher} 
\begin{equation}\label{eq:LO-C0}
C_0=1+\Delta_0^{(0)}(\rho)+
C_F\frac{\alpha_s}{\pi}\left\{\left(\frac{25}{8}-\frac{\pi^2}{2}\right)+
\Delta_0^{(1)}(\rho)\right\}
\end{equation}  
with $C_F=4/3$ and $\rho=m_c^2/m_b^2$. 
Here $\Delta_0^{(0)}(\rho)$ and $\Delta_0^{(1)}(\rho)$ are
corrections due to
charmed quark mass at LO and NLO respectively. They are  
known analytically and normalized such that 
$\Delta_0^{(0)}(0)=\Delta_0^{(1)}(0)=0$.

The coefficient $C_v$ is singled out by taking the matrix element
between quarks on shell and one gluon with vanishing momentum and
longitudinal polarization. The 
coefficient $C_v$ reads
\begin{equation}
C_v=5+C_F\frac{\alpha_s}{\pi}\left\{-\frac{25}{24}-\frac{\pi^2}{2}
\right\}\, .
\end{equation}
It has no $\mu$ dependence and no $C_A$ color contribution.
This matches also
the possibility to compute this coefficient using small momentum
expansion near the quark mass shell, $p=mv+k$. 
A powerful check of the result is an explicit cancellation of the
contribution proportional 
to the color structure $C_A$ and the renormalization (cancellation
of $\epsilon$-poles) with the same
renormalization constant~$Z_2^{OS}$ shown in~(\ref{ZOS}).

The final coefficient of the 
chromo-magnetic operator multiplied by~$C_m$ (see eq.~(\ref{HQEfin}))  
reads
\begin{equation}
C_{fin}=-C_v+(C_G - {\tilde C}_G C_0)/C_m
\end{equation}
and 
\begin{eqnarray}  \label{Cfin}
&&C_{fin}=-3+\Delta_G^{(0)}(m_c)+\frac{\alpha_s}{\pi}\Delta_G^{(1)}(m_c)\\
&&+\frac{\alpha_s}{\pi}
\left\{
C_A\left(\frac{31}{18}-\frac{\pi^2}{9}\right)+
C_F\left(\frac{43}{144}-\frac{19\pi^2}{36}\right) 
\right\} . \nonumber 
\end{eqnarray} 
The function $\Delta_G^{(0)}(\rho)$ is known 
analytically. The function $\Delta_G^{(1)}(\rho)$ 
emerges in the analysis of ref.~\cite{Gambino}
where the analytical result for the coefficient of
the chromo-magnetic operator at the level of hadronic structure
functions has been obtained. 
Both functions are chosen such that they vanish at $m_c=0$.
The final integration over the phase
space in ref.~\cite{Gambino} has been done numerically that prevents us
from a direct comparison between the two results.
Numerically we obtain at $m_c = 0$
\begin{eqnarray}\label{eq:NCfin}
C_{fin}&=&-3+\frac{\alpha_s}{\pi}
\left(
0.63 C_A-4.91C_F\right) \\
&=&-3+\frac{\alpha_s}{\pi}
\left(
-4.67\right)=-3(1+1.56\frac{\alpha_s}{\pi}) \nonumber   .
\end{eqnarray}
The $\mu$ dependence of  the prefactor of 
${\cal O}_G$ in~(\ref{HQE}) matches the leading order 
anomalous dimension of the chromo-magnetic operator~\cite{chromo-anom}, 
such that $C_{fin}$ is  $\mu$ independent. Furthermore, 
the mass parameter of the heavy quark  $m_b$ 
is chosen to be the pole mass 
which is a proper formal parameter for perturbative 
computations in HQET (see, discussion in~\cite{mannel-bigi}).
After having obtained the results of perturbation theory 
computation for the coefficients of HQE, one is free to change
this parameter to any other~\cite{again-kin}.

Our results~(\ref{Cfin},\ref{eq:NCfin}) still 
depend on $\mu$ through the strong coupling 
$\alpha_s$ defined in the $\overline{\rm MS}$-scheme; 
however, this remaining scale dependence can only be resolved at the 
next order in $\alpha_s $. 

\section{Discussion of the results}
\noindent
The radiative corrections are of reasonable magnitude and are well
under control for the numerical values of the coupling constant for 
$\mu\sim 2-4~{\rm GeV}$. This provides a clean application of the
results to decay into light quarks~$u$ for bottom mesons and 
$d$~for
charmed mesons.

For application to $b\to c$ transition 
the important question is the magnitude of corrections due to 
nonvanishing charmed quark mass.
It seems that mass corrections are important but still under control. 
The small $\rho$ expansion reads  
$\Delta_G^{(0)}(\rho) = 8\rho+\ldots$, and   
$\Delta_G^{(1)}(\rho) = A\rho+\ldots$   
where the factor $A$ is not known analytically.  
Assuming $|A|\leq 50$ one sees that  
the massless approximation dominates the radiative correction
for typical values of $\rho$ in the range  
$\rho=0.06\pm 0.02$~\cite{mc-mass}, 
\begin{equation}
C_{fin}=-3+\frac{\alpha_s}{\pi}
\left(
-4.67+\rho A\right) \, . 
\end{equation}
The numerical value of the coefficient can
change significantly only in the case of negative (and rather large)
contribution due to $c$-quark mass.

At present the value of $|V_{ub}|$ from inclusive decays is  
$|V_{ub}|=(4.41\pm 0.15\pm 0.16)\times 10^{-3}$~\cite{PDG} 
while the extraction from exclusive $B \to \pi \ell \bar{\nu}$ yields
$|V_{ub}|=(3.23\pm 0.31)\times 10^{-3}$. However, the exclusive 
determination does not rely on the local OPE considered here, so 
from our results we cannot really draw a definite
conclusion. Nevertheless, 
if 
our result indicates the size of the expected corrections, it cannot 
resolve the
tension between the inclusive and the exclusive value.  

More important are the implications for 
inclusive semileptonic $B$ meson decays to charm, since here  
the precision is high enough to worry about the correction computed above. 
Indeed, the inclusive determination has a precision 
at the level of roughly 2\%,  
the value being 
$|V_{cb}|=(42.4\pm 0.9)\times 10^{-3}$~\cite{Gambino:2013rza,PDG}. 
Since we only have the analytical result in the limit $m_c \to 0$ at hand, 
we estimate the impact of our correction in a simplified manner.
Because it is a small correction, we only account for charmed quark 
mass at tree approximation, taking into account the 
kinematic function
$
\Delta_0^{(0)}(\rho) = -8\rho-12\rho^2\ln \rho+8\rho^3-\rho^4 
$.
The determination of $|V_{cb}|$ uses 
the total rate only, so we get for the shift in $|V_{cb}|$
through the  $\alpha_s$ correction in the coefficient of the
chromo-magnetic operator 
\begin{equation}
\frac{\Delta |V_{cb}|}{|V_{cb}|} = 4.67\frac{\alpha_s}{\pi}
\frac{3\Delta m_B^2}{8m_b^2} \frac{1}{2(1+\Delta_0^{(0)}(\rho))}
\end{equation} 
which yields for $\rho = 0.07$ and $\alpha_s/\pi=0.1$ 
a relative shift of $+0.3\%$
in the value of $|V_{cb}|$. This result is compatible 
with the study in~\cite{Gambino}, 
which includes the charmed 
quark mass. A preliminary comparison of extrapolation of the
  results of ref.~\cite{Gambino} to small mass 
limit shows a reasonable agreement.  

The shift in  $|V_{cb}|$ has to be compared to the  
corrections of order $(\Lambda/m_b)^n$, $n=3,4$ 
at tree level. The $(\Lambda/m_b)^3$ 
contributions induce a relative shift in $|V_{cb}|$ 
of about $-1.5\%$ which is included in the current analysis. 
The terms of order 
$(\Lambda/m_b)^4$ are not yet included and shift the value 
of $|V_{cb}|$ by about 
$0.3\%$~\cite{HigherOrders}, which is roughly 
of the same order as the corrections considered here.

\begin{acknowledgments}
We acknowledge interesting 
discussions with A.G.~Grozin, B.O.~Lange, J.~Heinonen, and 
T.~Huber.
We thank P.~Gambino for a careful reading of the manuscript, valuable
comments, and discussion of the 
results of the small charmed mass extrapolation.
This work is supported by  
DFG Research Unit FOR 1873 ``Quark Flavors Physics
and Effective Theories''.
\end{acknowledgments}

\end{document}